\documentclass[aps,prx,groupedaddress,twocolumn,showpacs,longbibliography,10pt]{revtex4-2}

\usepackage[table]{xcolor}
\usepackage{amsmath}
\usepackage{slashed}
\usepackage{amssymb}
\usepackage{footmisc}

\usepackage{graphicx}
\usepackage{draft}
\usepackage{color}
\usepackage{times, verbatim}      
\usepackage{mathtools,tikz}

\usetikzlibrary{arrows.meta,calc,decorations.pathmorphing,positioning}

\usepackage{pgfplots}
\usetikzlibrary{decorations.pathmorphing}
\usetikzlibrary{matrix}
\usepackage{natbib}
\usepackage{multirow}
\usepackage{hyperref}

\definecolor{colour1}{rgb}{0.368417, 0.506779, 0.709798}
\definecolor{colour2}{rgb}{0.880722, 0.611041, 0.142051}
\definecolor{colour3}{rgb}{0,1,1}
\definecolor{colour4}{rgb}{0,1,0}
\definecolor{colour5}{rgb}{1,1,0}
\usetikzlibrary{decorations.pathreplacing}
\usepackage{dsfont}

\usepackage{amsthm}

\hypersetup{
       colorlinks=true,
}

\usepackage{multirow}
\makeatletter
\newsavebox{\@brx}
\newcommand{\llangle}[1][]{\savebox{\@brx}{\(\m@th{#1\langle}\)}%
  \mathopen{\copy\@brx\kern-0.5\wd\@brx\usebox{\@brx}}}
\newcommand{\rrangle}[1][]{\savebox{\@brx}{\(\m@th{#1\rangle}\)}%
  \mathclose{\copy\@brx\kern-0.5\wd\@brx\usebox{\@brx}}}
\makeatother

\newcommand{\ep}{\epsilon}

\newcommand{\pa}{\partial}

\newcommand{\SO}{{\text{SO}}}

\newcommand{\lap}{\nabla^2}

\newcommand{\GN}{\mathrm{GN}}
\newcommand{\Li}{\operatorname{Li}_2}
\newcommand{\atanh}{\operatorname{artanh}}

\begin{document}
 
\author{Oleksandr Diatlyk$^{1}$, Zimo Sun$^{2,3}$, and Yifan Wang$^{1,2}$
}
\affiliation{$^{1}$ Center for Cosmology and Particle Physics, New York University, New York, NY 10003, USA}
\affiliation{$^{2}$ Institute for Advanced Study, Princeton, NJ 08540, USA}
\affiliation{$^{3}$ Joseph Henry Laboratories, Princeton University, Princeton, NJ 08544, USA}

\title{ \hfill{\normalfont\small PUPT-2665}\\[2em]
Defect Anomalies at a Distance
}

\begin{abstract}
Surface defects carry Weyl and gravitational anomalies that generalize the central charges of two-dimensional conformal field theories (CFTs) and characterize the anomalous response of gapless surface modes to background geometry. These anomalies usually appear as contact terms on the defect. We explain how they are encoded at a distance, in scheme-independent integrated bulk stress-tensor correlators. We derive sum rules for them in terms of defect operator data, establish a lower bound on the boundary Weyl anomaly in three-dimensional CFTs, and illustrate our results with explicit examples.
\end{abstract}

\date{\today}


\maketitle

\section{Introduction and Summary}

Conformal central charges provide a universal measure of gapless degrees of freedom in two spacetime dimensions. In a two-dimensional conformal field theory (CFT), the left- and right-moving sectors carry central charges $c_L$ and $c_R$, which fix the vacuum stress-tensor correlators on the plane~\cite{Belavin:1984vu}. They are also anomalies in a background geometry: $c=(c_L+c_R)/2$ sets the response to a local rescaling of the metric~\cite{Polyakov:1981rd}, and the chiral central charge $\kappa=c_L-c_R$ detects the violation of diffeomorphism invariance~\cite{Alvarez-Gaume:1983ihn}. They govern universal observables, including the Casimir energy and momentum on a circle, thermal and entanglement entropies, and the asymptotic density of states~\cite{Bloete:1986qm,Affleck:1986bv,Holzhey:1994we,Calabrese:2004eu,Cardy:1986ie}. Under unitary Poincar\'e invariant renormalization group (RG) flows, $c$ decreases between fixed points~\cite{Zamolodchikov:1986gt}, whereas $\kappa$ is invariant~\cite{tHooft:1979rat}.

Rather than existing in isolation, such gapless modes often arise on a surface in a higher-dimensional system. Familiar examples are chiral fermions localized on domain walls or vortices of massive fermions~\cite{Jackiw:1975fn,Jackiw:1981ee,Callan:1984sa}, where the bulk is gapped and the surface is a standalone 2d CFT at long distances (coupled to a topological theory in higher dimensions). A gapless bulk allows much more \cite{Bray:1977fvl,Diehl:1996kd,Giombi:2020rmc,Metlitski:2020cqy,Padayasi:2021sik,Toldin:2021kun,Krishnan:2023cff,Trepanier:2023tvb,Giombi:2023dqs,Raviv-Moshe:2023yvq,Cuomo:2023qvp,Diatlyk:2024ngd,Popov:2025cha,Giombi:2021cnr,Herzog:2022jlx,Jiang:2025sfb,Fedorenko:2026svv,Diatlyk:2026gab}: the surface exchanges energy and momentum with the bulk, its effective dynamics is nonlocal, and it need not have a conserved stress tensor of its own. Its Weyl and gravitational anomalies remain well defined, with coefficients $b$ and $\kappa$ that generalize the central charges, and for RG flows driven by local surface perturbations with the bulk at a fixed point, $b$ decreases and $\kappa$ is invariant~\cite{Jensen:2015swa,Shachar:2022fqk,Casini:2023kyj,Wang:2020xkc}.

These parallels raise a basic question: how are the surface anomalies encoded in local correlation functions? We answer it for two-dimensional conformal defects by connecting their Weyl and gravitational anomalies to correlators of the bulk stress tensor.

Consider a conformal surface defect on $\Sigma$: a boundary or interface of a 3d CFT~\cite{McAvity:1995zd}, or a surface operator in higher dimensions~\cite{Gukov:2006jk,Drukker:2020dcz}. We use coordinates $x^\m=(z,\bar z,y^i)$, with $\Sigma$ at $y^i=0$, $r\equiv|y|$ the distance from $\Sigma$ and $z=x^1+ix^2$. We define the defect contribution to the generating functional and the stress tensor by
\ie
W[g]\equiv-\log{Z_\Sigma[g]\over Z[g]}\,,\quad\la T^{\m\n}\ra=-{2\over\sqrt g}{\D W[g]\over\D g_{\m\n}}\,.
\label{Tdef}
\fe
The trace of $T$ is a contact term that detects the Weyl anomaly~\cite{Graham:1999pm,Henningson:1999xi,Jensen:2018rxu,Chalabi:2020iie},
\ie
\la T^\m_\m\ra=\cA\D_\Sigma\,,~\cA={1\over24\pi}\left(bR_\Sigma+d_1\mathring\Pi^2-d_2W_{ab}{}^{ab}\right)\,,
\label{Weylanomaly}
\fe
with $\D_\Sigma$ the delta function on $\Sigma$, $\mathring\Pi$ the traceless second fundamental form (extrinsic curvature tensor) and $W_{ab}{}^{ab}$ the tangential contraction of the ambient Weyl tensor. The coefficients $d_1,d_2$ are fixed by flat-space data~\cite{Bianchi:2015liz,Herzog:2017kkj,Jensen:2018rxu} 
\ie
d_1={3\pi^2\over4}c_D\,,~~d_2={6\pi V(d-1)\over d-3}h\,,
\label{d1d2}
\fe
where $V$ is the volume of ${\rm S}^{d-3}$.

For flat $\Sigma$ in $\mR^d$, $c_D$ governs the displacement operator two-point function~\cite{Billo:2016cpy}
\ie
\pa_\m T^{\m i} = -\D_\Sigma D^i\,,\quad
\la D^i(z,\bar z) D^j(0)\ra ={c_D \D^{ij}\over |z|^6}\,,
\label{displacement}
\fe
and $h$ determines the stress tensor one-point function (which vanishes for $d=3$)
\ie
\la T_{ab}\ra=-h {\D_{ab}\over r^d}\,,~ \la T_{ai}\ra=0\,,~ \la T_{ij}\ra={h(3 \D_{ij}-d {y_i y_j\over r^2})\over (d-3)r^d}\,.
\label{Tonepoint}
\fe
The coefficients $c_D$ and $h$ are tied to the trace anomaly \eqref{Weylanomaly} through the scale dependence of the renormalized action $W[g]$: the correlators in~\eqref{displacement} and~\eqref{Tonepoint} have non-integrable singularities, $|z|^{-6}$ on $\Sigma$ and $r^{-d}$ at $r\to0$, whose distributional extensions introduce a scale. Their separated-point correlator coefficients therefore also determine the corresponding scale anomalies. 
The coefficient $b$ in \eqref{Weylanomaly}, normalized to $b=1$ for a free boson on $\Sigma$, is not captured this way. Indeed, under a constant rescaling, $\m\pa_\m W[g]=-\int_\Sigma\sqrt\gamma\,\cA$. While the integrals of $\mathring\Pi^2$ and $W_{ab}{}^{ab}$ respond to local variations of the background geometry,  $\int_\Sigma\sqrt\gamma R_\Sigma$ is topological and has vanishing local variation. 

On a planar defect, $b$ is instead detected through a Weyl rescaling that varies along $\Sigma$, equivalently encoded in a defect-supported contact term in the flat-space correlator $\la T^\m_\m T_{ab}\ra$. Here we will extract it from an integrated stress-tensor correlator in the bulk.
A defect that breaks parity along $\Sigma$ can also carry a gravitational anomaly \cite{Wang:2020xkc}, whose coefficient is normalized to $\kappa=c_L-c_R$ for a decoupled 2d CFT. It too appears as a contact term in $\la TT\ra$, through the Ward identity for diffeomorphism.

General contact terms are subject to scheme dependence due to finite local counterterms, whereas anomaly coefficients and separated-point correlators are scheme independent. We therefore seek a characterization of the anomalies in terms of separated-point data, analogous to $\la T_{zz}T_{zz}\ra=c_L/(8\pi^2z^4)$ in a 2d CFT. There, stress-tensor conservation relates the anomalous contact term to the separated-point correlator in a simple manner \cite{DiFrancesco:1997nk}.

For a defect, the reduced conformal symmetry leaves $\la TT\ra$ dependent on conformal cross-ratios in a nontrivial way, so there is no single coefficient to read-off. Instead, as we show, Ward identities relate the traceless tangential part of the contact term in $\la T^\m_\m T_{ab}\ra$ to an integrated stress-tensor correlator which is scheme independent. Such a contact term receives contributions from both the $R_\Sigma$ and $W_{ab}{}^{ab}$ terms in \eqref{Weylanomaly} with the overall coefficient 
\ie
B=b-6\pi hV\,,
\label{Banom}
\fe
which also governs the logarithmic contribution to the defect entanglement entropy of a ball $S_{\rm def}=B\log(L/\ep)/3$ \cite{Jensen:2018rxu,Kobayashi:2018lil}. Since $h$ is independently determined by~\eqref{Tonepoint}, an expression for $B$ in terms of $\la TT\ra$ also fixes $b$. A similar analysis based on the Ward identities fixes $\kappa$ in terms of another integral of $\la TT\ra$. The main results here are,  
\ie
B=&\,12\pi Vr^d\,{\rm RV}\int d^dx'\log\left(r\over r'\right)\\
&\times\la T_{zz}(x)T_{\bar z\bar z}(x')+(x\leftrightarrow x')\ra\,,\\
\kappa=&\,24\pi Vr^d\,{\rm RV}\int d^dx'\log\left(r\over r'\right)\\
&\times\la T_{\bar z\bar z}(x)T_{zz}(x')-(x\leftrightarrow x')\ra\,,
\label{Bkappa}
\fe
for any $x$ away from $\Sigma$. Here $\la TT\ra$ denotes the connected correlator with the defect (normalized by subtracting the two-point function without defect), and ${\rm RV}$ is minimal subtraction of the power-law singularities at $x'\to x$ (see end of Section~\ref{sec:transport}).

We emphasize that \eqref{Bkappa} is insensitive to finite local counterterms: c-number bulk counterterms cancel between $Z_\Sigma$ and $Z$, while defect counterterms have no support at $x$ \footnote{Certain contact terms at $x'=x$ nevertheless contribute. They arise from bulk scalars of dimension $d-2$ with nonzero one-point functions, the term $A_{d-2}$ of~\eqref{eq:sumrule}, and, for $\kappa$ in $d=3$, from a pseudoscalar, already included in that sum. Their coefficients are fixed by conservation and tracelessness of $T_{\m\n}$ (end of Section~\ref{sec:transport}) and thus scheme independent. Other operator-valued counterterms, such as $\int\!\sqrt g\,W^2O$ with $\Delta_O=d-4$ for $d\ge6$, do not change~\eqref{Bkappa}.}.

The structures of these identities become more transparent after a Weyl map to $\mH^3\times{\rm S}^{d-3}$, where the preserved symmetry $\SO(3,1)\times\SO(d-2)$ acts by isometries and $\Sigma$ lies at the boundary $r=0$
\ie
ds^2=dz\,d\bar z+dy^2=r^2\left({dr^2+dz\,d\bar z\over r^2}+d\Omega_{d-3}^2\right)\,.
\label{Weyl}
\fe
Writing $X=(z,\bar z,r)$ and $G_{abcd}(x,x')=\la T_{ab}(x)T_{cd}(x')\ra$ for correlators of the stress tensor in its tangential components to $\Sigma$, we define the sphere averaged correlator on $\mH^3$ by
\ie
F_{abcd}(X,X')\equiv(rr')^dV^{-2}\int d\Omega\,d\Omega'\,G_{abcd}(x,x')\,.
\label{FH3}
\fe
Here the tangential components of $F$ are expressed in the frame $(dz,d\bar z)/r$. The identities~\eqref{Bkappa} then become
\ie
B&=-3\pi V^2\,{\rm RV}\int_{\mH^3}d{\rm vol}'\log{r'\over r}\,F_{(ab)(ab)}(X,X')\,,\\
\kappa&=6\pi i\,V^2\,{\rm RV}\int_{\mH^3}d{\rm vol}'\log{r'\over r}\,\ep_{ac}F_{(ab)(bc)}(X,X')\,,
\label{BH3}
\fe
where $d{\rm vol}'=dr'd^2z'/r'^3$ is the $\mH^3$ volume form and $\ep_{12}=1$ is the antisymmetric tensor. Explicitly $F_{(ab)(ab)}=4(F_{zz\bar z\bar z}+F_{\bar z\bar zzz})$ and $\ep_{ac}F_{(ab)(bc)}=-4i(F_{zz\bar z\bar z}-F_{\bar z\bar zzz})$. 

Using the $\mH^3$ isometry, we set $X=(0,1)$ and $X'=(z',e^\sigma)$. Integrating over the horosphere at constant $\sigma$ gives
\ie
B&=-3\pi V^2\,{\rm RV}\int_{-\infty}^\infty d\sigma\,\sigma\,F_+(\sigma)\,,\\
\kappa&=6\pi V^2\,{\rm RV}\int_{-\infty}^\infty d\sigma\,\sigma\,F_-(\sigma)\,,
\label{Bsingleintegral}
\fe
with
\ie
F_\pm(\sigma)\equiv4e^{-2\sigma}\int d^2z'\,\big(F_{zz\bar z\bar z}\pm F_{\bar z\bar zzz}\big)(X,X')\,.
\label{Fpm}
\fe
In a standalone 2d CFT, the stress tensor is conserved on the surface, so the mixed correlators in~\eqref{Fpm} are purely contact terms and vanish at separated points. Coupling to a gapless bulk relaxes this conservation law, and allows these correlators to be nonzero at separated points. Their first moments in $\sigma$ give the surface anomalies \eqref{Bsingleintegral}.
 
The rest of the paper is organized as follows.
We review the defect Ward identities in Section~\ref{sec:ward} and use them in Section~\ref{sec:transport} to transport the Weyl anomaly to correlation functions at separated points, which proves the identity for $B$ in \eqref{Bkappa}. Section~\ref{sec:chiral} extends this relation to $\kappa$. Section~\ref{sec:spectrum} gives sum rules for the surface anomalies over the defect spectrum. Section~\ref{sec:examples} applies the results to boundaries in three dimensions, where they bound $b$ from below by $c_D$, and to  monodromy defects for free fields in four dimensions.  

\section{Defect Ward Identities}\label{sec:ward}

Let us first review the defect Ward identities \cite{Billo:2016cpy} while taking into account the defect anomalies, which are key to deriving \eqref{Bkappa}. We choose the scheme where the parity-odd part of $W$ is Weyl invariant. Then under a diffeomorphism that maps $\Sigma$ to itself the effective action changes only by the gravitational anomaly, and under a Weyl rescaling only by the Weyl anomaly,
\ie
W[g+\cL_\xi g]&=W[g]-\kappa\int_\Sigma\sqrt\gamma\,\xi^a\cA_a+\cO(\xi^2)\,,\\
W[ge^{2\varphi}]&=W[g]-\int_\Sigma\sqrt\gamma\,\varphi\,\cA+\cO(\varphi^2)\,.
\label{WWeyl}
\fe
Here $\cA_a$ is a local function of the induced metric, normalized so that $\kappa=c_L-c_R$ for a decoupled 2d CFT (see Section~\ref{sec:chiral}).  

Varying $g$ in \eqref{WWeyl}, we obtain
\ie
\D_\xi\la T_{\m\n}(x)\ra_g=&\,\cL_\xi\la T_{\m\n}(x)\ra_g+\int_\Sigma\sqrt\gamma\,\xi_i\la D^iT_{\m\n}(x)\ra_g\\
&-2\kappa\,{\D\over\D g^{\m\n}(x)}\int_\Sigma\sqrt\gamma\,\xi^a\cA_a\,,\\
\D_\varphi\la T_{\m\n}(x)\ra_g=&\,(2-d)\,\varphi\,\la T_{\m\n}(x)\ra_g\\
&-2\,{\D\over\D g^{\m\n}(x)}\int_\Sigma\sqrt\gamma\,\varphi\,\cA\,.
\label{Ward}
\fe
Here $T_{\m\n}$ is the full stress tensor as in \eqref{Tdef} and contains the bulk stress tensor $T_{\rm b}^{\m\n}$ and defect operators such as $\hat B^{ab}$ and $\hat C^{abi}$, which couple to the induced metric and extrinsic curvature of $\Sigma$ respectively. A metric perturbation $\D g_{ab}=s_{ab}$ in components tangential to $\Sigma$  gives
\ie
\D_s\la T_{\m\n}\ra_g={}&{1\over2}\int s_{ab}\la T_{\rm b}^{ab}T_{\m\n}\ra_g+{1\over2}\int_\Sigma s_{ab}\la\hat B^{ab}T_{\m\n}\ra_g\\
&-{1\over4}\int_\Sigma\pa_is_{ab}\la\hat C^{abi}T_{\m\n}\ra_g\,.
\label{Texpvar}
\fe
The first term is where $\la TT\ra$ enters. Since $W$ is renormalized, these integrated correlators are regularized by RV (see details near the end of Section~\ref{sec:transport}), and finite counterterms shift the contact terms of
\ie
G_{\m\n\rho\sigma}(x,x')=-{4\over\sqrt{g(x)g(x')}}{\D^2W\over\D g^{\m\n}(x)\,\D g^{\rho\sigma}(x')}\,.
\label{conv}
\fe
Neither the anomalies nor $\la TT\ra$ at separated points depend on these counterterms, and for suitable $\xi$ and $\varphi$ the Ward identities relate the two as we will show below.

\section{Transporting Weyl Anomalies to the Bulk}\label{sec:transport}

The task is to transport the anomaly, a contact term on $\Sigma$, to correlators of $T$ away from $\Sigma$. We start with the Weyl anomaly and assume $\kappa=0$. The response to a general metric perturbation is
\ie
\D\la T_{\m\n}(x)\ra_g={}&{1\over2}\int d^dx'\sqrt{g(x')}\,G_{\m\n}{}^{\rho\sigma}(x,x')\,\D g_{\rho\sigma}(x')\\
&-{1\over2}\D g^\rho{}_\rho(x)\la T_{\m\n}(x)\ra_g\,,
\label{response}
\fe
where the last term comes from the $\sqrt g$ in~\eqref{Tdef} and vanishes for a traceless source.
An infinitesimal Weyl transformation $\D g=2\varphi g$ couples only to the trace $T^\m{}_\m(x')$, which vanishes unless $x'=x$ or $x'\in\Sigma$, reducing to the second line of \eqref{Ward}. A diffeomorphism instead transports operator insertions, and it changes the metric perturbation but not the anomaly. We combine the two into a deformation $\D g=\cL_\xi g+2\varphi g$ with three properties. It has only tangent components, $\D g_{ab}=s_{ab}$, so \eqref{Texpvar} applies where only $T_{ab}(x')$ is integrated. It is traceless, $s_{aa}=0$, so its response in \eqref{Texpvar} no longer contains the contact terms from \eqref{Weylanomaly}. It vanishes on $\Sigma$ with its first derivatives, so the metric and the extrinsic curvature of $\Sigma$ do not change, and by~\eqref{Texpvar} no defect operators are triggered. 

The Ward identities \eqref{Ward} then determine the linear response of this $\D g$ by the Weyl anomaly, up to the Lie transport of $\la T\ra$, while \eqref{Texpvar} writes the same response as an integral of $\la TT\ra$. 
We choose the un-integrated stress tensor to be in the tangential directions so that these responses are non-vanishing. Then equating the two produces an identity for $B$ in terms of the integral of $\la T_{ab}T_{cd}\ra$. We take $\Sigma$ to carry no decoupled 2d CFT; these integrals then converge for a unitary defect with a discrete spectrum (Section~\ref{sec:spectrum}).

It remains to identify the pair of diffeomorphism $\xi$ and Weyl factor $\varphi$ that satisfy the three properties laid out above. 
This is achieved by extending a conformal map $u(z)$ with conformal factor $\omega=-{\rm Re}\,\pa_z u$ on $\Sigma$ into the bulk $\mR^d$ as follows,
\ie
\xi=\Bigl[\bigl(u-{1\over2}r^2 \pa^2_{\bar z}\bar u \bigr)\pa_z+{\rm c.c.}\Bigr]-\omega\,y^i\pa_i\,,
\label{xi}
\fe
and $\varphi=\omega$ stays independent of the transverse directions to $\Sigma$. This gives a tangential metric source $s_{ab}$ whose only nonzero components are
\ie
s_{zz}=-{1\over2}r^2\pa_z^3u\,,~s_{\bar z\bar z}=-{1\over2}r^2\pa_{\bar z}^3\bar u\,.
\label{lift}
\fe
Note that the vector field \eqref{xi} and metric perturbation \eqref{lift} have a natural origin in the $\mH^3$ frame (see \eqref{Weyl}): since $\xi^r=-\omega r$, we have $\cL_\xi g_{\mH^3}=r^{-2}s_{ab}dx^adx^b$, so $\xi$ is the Brown-Henneaux vector field that lifts a 2d conformal map $u(z)$ on the boundary of $\mH^3$ into the bulk and induces a metric perturbation in the Fefferman-Graham gauge which corresponds to boundary graviton excitations \cite{Brown:1986nw,Imbimbo:1999bj,Du:2024tlu}. 

Now there are infinitely many Ward identities one can derive upon choosing the conformal map $u(z)$. Here we choose the simplest one such that \eqref{lift} is nonzero, 
\ie 
u=-{z^3\over 3}\,,~ s_{zz}=s_{\bar z\bar z}=r^2\,,~\omega={\rm Re}\,z^2\,.
\label{choiceu}
\fe 

Let us inspect the variation of $T_{\bar z\bar z}(x)$ under this metric perturbation. In \eqref{Texpvar}, 
the source $s_{zz}$ does not contribute since $\la T_{\bar z\bar z}(x')T_{\bar z\bar z}(x)\ra$ has spin $4$ under rotations along $\Sigma$ about $x$ and thus integrates to zero. In \eqref{Ward} the displacement contribution drops out since $\xi^i=0$ on $\Sigma$ 
and $(2-d)\omega\la T_{\bar z\bar z}\ra$ vanishes by \eqref{Tonepoint}. The anomaly terms contribute $\D(\sqrt\gamma R_\Sigma)=4\pa_{\bar z}\pa_{\bar z}\D g_{zz}$ and $\D W_{ab}{}^{ab}={4(d-3)\over d-1}\pa_{\bar z}\pa_{\bar z}\D g_{zz}$, while $\mathring\Pi^2$ is quadratic in the metric variation, producing altogether 
\ie 
\D\int_\Sigma\sqrt\gamma\,\omega\,\cA={B\over 6\pi}\int_\Sigma\D g_{zz}\pa_{\bar z}\pa_{\bar z}\omega\,.
\fe
Here $\pa_{\bar z}^2\omega=1$ from \eqref{choiceu}. Then putting together the defect Ward identities \eqref{Ward}, we obtain in flat space,
\ie
2\int d^dx'\,r'^2\la T_{zz}(x')T_{\bar z\bar z}(x)\ra=\cL_\xi\la T_{\bar z\bar z}(x)\ra+{B\over12\pi}\D_\Sigma(x)\,,
\label{W1}
\fe
where the Lie derivative of the one-point function \eqref{Tonepoint} is
\ie
\cL_\xi\la T_{\bar z\bar z}(x)\ra=2\pa_{\bar z}\xi^z\la T_{z\bar z}(x)\ra=-h\,r^{2-d}\,.
\fe
It is useful to rewrite \eqref{W1} using another Ward identity from \eqref{Ward} with no Weyl rescaling but a linear diffeomorphism $\xi_0=\bar z\pa_z+z\pa_{\bar z}$. The similar steps then give
\ie
2\int d^dx'\,\la T_{zz}(x')T_{\bar z\bar z}(x)\ra=\cL_{\xi_0}\la T_{\bar z\bar z}(x)\ra=-h\,r^{-d}\,.
\label{W2}
\fe
Unlike~\eqref{lift}, this constant source does not vanish on $\Sigma$, but it couples only to the traceless part of $\hat B^{ab}$, of dimension $2$: a primary would be a conserved stress tensor on $\Sigma$, absent here, and a descendant integrates to zero against a weight independent of $z'$.

Putting \eqref{W1} and \eqref{W2} together, we find
\ie
\int d^dx'\,(r'^2-r^2)\la T_{zz}(x')T_{\bar z\bar z}(x)\ra={B\over24\pi}\D_\Sigma(x)\,,
\label{eqanom}
\fe
where the LHS is understood to be a distribution in $x$. The transports cancel as distributions: with $\la T_{ab}\ra=\D_{ab}t$ they are $r^2t$ in~\eqref{W1} and $t$ in~\eqref{W2}, even if $t$ has contact terms on $\Sigma$. The LHS also contains contact terms on $\Sigma$, which are scheme dependent but shift its two terms equally and leave $B$ fixed. For example, $\lambda\int_\Sigma\!\sqrt\gamma\,R$, with $R$ the bulk curvature, shifts $\int d^dx'\,r'^2G$ by $2\lambda(d-2)\D_\Sigma$ and $\int d^dx'\,G$ by $\lambda\pa_y^2\D_\Sigma$, and $r^2\pa_y^2\D_\Sigma=2(d-2)\D_\Sigma$.

To isolate an identity for $B$, we integrate against a smooth test function $f(r)$ with compact support and $f=1$ near $r=0$, which gives
\ie
{B\over24\pi}=\int\! d^{d-2}y'\!\int\! d^dx\,\bigl[f(r)-f(r')\bigr]r'^2\la T_{zz}(x')T_{\bar z\bar z}(x)\ra\,.
\label{smeared}
\fe
Here we have used the identity
\ie 
\int d^{d-2}y \int d^d x' (f(r)r^2-f(r')r'^2) \la T_{zz}(x')T_{\bar z\bar z}(x)\ra =0\,,
\label{parityid}
\fe
which follows from \eqref{W2} applied at $x$ and its image under longitudinal parity $z\leftrightarrow\bar z$ and applied at $x'$, together with translation symmetry along $\Sigma$. The integration kernel  in \eqref{smeared} vanishes where both points are near $\Sigma$, so neither the contact terms on $\Sigma$ nor the scale anomalies of $G$ supported there contribute. Using overall dilation and transverse rotation symmetry, we can factor out the integral over $r'$ at fixed $\sigma=\log(r/r')$ in \eqref{smeared},
\ie
\int_0^\infty{dr'\over r'}\,\bigl[f(r'e^\sigma)-f(r')\bigr]=-\sigma\,.
\fe
Consequently, after relabeling the two points,
\ie
B=24\pi Vr^d\,{\rm RV}\int d^dx'\log\left(r\over r'\right)\la T_{zz}(x)T_{\bar z\bar z}(x')\ra\,.
\label{Bflat}
\fe
Had we chosen to study the variation of $T_{zz}(x)$ in \eqref{Ward}, we would obtain \eqref{Bflat} with $z$ and $\bar z$ exchanged, and they both equal to the first line of
\eqref{Bkappa} when longitudinal parity is preserved.

Let us comment on the regularization needed to define the integral \eqref{Bflat}. Near $x'=x$ the integral has only power divergences and no logarithmic ones, and RV is their minimal subtraction by analytic continuation. Indeed, the identity channel in the bulk OPE cancels in $\la TT\ra=\la TT\ra_\Sigma-\la TT\ra_0$, and by unitarity only $T$ and scalars with $\la O\ra\neq0$ can still contribute a non-integrable divergence \cite{Billo:2016cpy}. The integration weight in \eqref{Bflat} vanishes at $x'=x$ and removes the potential $T$ contribution. A bulk scalar $O$ appearing in the $TT$ OPE contributes $\Pi(\pa)|x-x'|^{\Delta_O-2d+4}\la O\ra$ and similarly for its descendants, where $\Pi(\pa)$ is fixed by conservation and tracelessness~\cite{Osborn:1993cr},
\ie
\Pi(\pa)={d-2\over d-1}\pa_z^2\pa_{\bar z}^2-\pa_z\pa_{\bar z}\Box+{1\over4}\Box^2\,,\quad\Box=4\pa_z\pa_{\bar z}+\pa_y^2\,.
\label{Pi}
\fe
The kernel in \eqref{Bflat} makes this integrable. It is smooth at $x'=x$, so integration by parts moves the derivatives onto it, and it does not depend on $z'$, so $\Pi(\pa)$ acts on it simply as ${1\over4}\Box^2$, giving ${(d-4)(d-6)\over 2r'^4}$. The reduced singularity $|x-x'|^{\Delta_O-2d+4}$ is integrable for $\Delta_O>d-4$, and this is automatic for $d<6$ by unitarity. With a hard-cutoff at $|x-x'|=\ep$, the power law divergences are the boundary terms from this integration by parts, which the analytic continuation in $\Delta_O$ subtracts automatically. Similar regularization for integrated correlators has been considered in \cite{Friedan:2012hi,Balthazar:2022hzb}.

Extra care is needed for $\Delta_O=d-2$, the dimension of a scalar that can enter the improvement of $T$~\cite{Callan:1970ze}. In this case $|x-x'|^{2-d}$ is harmonic, so the bulk Laplacian $\Box$ in~\eqref{Pi} gives a contact term $\la O\ra\pa^2\D^d(x-x')$ in the $TT$ OPE. It leaves a finite term upon regularization, since the $\log(r'/r)$ kernel vanishes at $x'=x$ but its derivatives do not. A priori this contact term can be tuned by the counterterm $\zeta\int\!\sqrt g\,RO$ (e.g. the familiar conformal mass of a free scalar), which also tunes the improvement of $T$. The tracelessness of $T$ fixes $\zeta$, so the contact term is scheme independent. It gives the term $A_{d-2}$ in~\eqref{eq:sumrule}, which is essential for the free scalar example of Section~\ref{sec:monodromy}. 

\section{Surface Gravitational Anomalies}\label{sec:chiral}

We now restore the gravitational anomaly $\kappa$ and follow Section~\ref{sec:transport} to see what changes. The key difference is that $\la TT\ra$ is no longer parity invariant for $\kappa\neq 0$, and its parity-odd part is tied to the anomaly through the first Ward identity in \eqref{Ward}. The anomaly is the non-conservation of the covariant stress tensor $\tilde T$ on $\Sigma$~\cite{Alvarez-Gaume:1983ihn,Bertlmann:2000da,Kraus:2005zm,Wang:2020xkc},
\ie
\nabla_\m\la\tilde T^{\m a}\ra={i\kappa\over96\pi}\,\D_\Sigma\,\ep^{ab}\pa_bR_\Sigma\,.
\label{covcons}
\fe
In the scheme of \eqref{WWeyl}, $\tilde T$ differs from $T$ in \eqref{Tdef} by a traceless local background-dependent Bardeen-Zumino term \cite{Bardeen:1984pm}. This term and~\eqref{covcons} are both supported on $\Sigma$, so the anomaly sees $\xi$ only through its pullback $U=\xi^a\pa_a|_\Sigma$, which for~\eqref{xi} is the conformal map $u(z)$. Since $\cL_\xi g$ pulls back to $\cL_U\gamma=-2\omega\gamma$, which is pure trace, the traceless term drops out of $\D_\xi W=-{1\over2}\int\sqrt g\,\la T^{\m\n}\ra\cL_\xi g_{\m\n}$, and here we may replace $T$ by $\tilde T$. Integrating~\eqref{covcons} against $\xi$ then gives the anomalous change of $W$ in~\eqref{WWeyl}, ${i\kappa\over48\pi}\int_\Sigma\theta R_\Sigma$, where $\theta={1\over2}\ep^{ab}\pa_aU_b$ parametrizes a frame rotation on $\Sigma$.  

For a conformal map on $\Sigma$ as in Section~\ref{sec:transport}, $\theta={\rm Im}\,\pa_zu$ and the Weyl factor and frame rotation combine to $\omega-i\theta=-\pa_zu$. The total anomalous variation of $W$ in \eqref{WWeyl} in this near flat background is then
\ie
&{1\over48\pi}\int_\Sigma\Bigl[\bigl(B+{\kappa\over2}\bigr)\pa_zu+\bigl(B-{\kappa\over2}\bigr)\pa_{\bar z}\bar u\Bigr]R_\Sigma\\
&={1\over12\pi}\int_\Sigma\Bigl[\bigl(B+{\kappa\over2}\bigr)\D\gamma_{\bar z\bar z}\pa_z^3u+\bigl(B-{\kappa\over2}\bigr)\D\gamma_{zz}\pa_{\bar z}^3\bar u\Bigr]\,.
\label{eq:total}
\fe 
One recognizes that here $B+\kappa/2$ and $B-\kappa/2$ enter in the same way as the $c_L$ and $c_R$ central charges of a decoupled 2d CFT \footnote{The only local parity-odd counterterm at this order, $i\int_\Sigma(\D\gamma_{\bar z\bar z}\pa_z^2\D\gamma^a{}_a-\D\gamma_{zz}\pa_{\bar z}^2\D\gamma^a{}_a)$, moves anomaly between the two lines of \eqref{WWeyl} but leaves \eqref{eq:total} unchanged.}. 

Since $\D\gamma_{zz}$ couples to $T_{\bar z\bar z}$, $B$ becomes $B-\kappa/2$ in~\eqref{W1} and~\eqref{Bflat}, we thus arrive at the following identity,
\ie 
B-{\kappa\over2}=&\,24\pi Vr^d\,{\rm RV}\int d^dx'\log\left(r\over r'\right)\la T_{zz}(x)T_{\bar z\bar z}(x')\ra\,,
\fe 
and its image under longitudinal parity is now an independent identity,
\ie
B+{\kappa\over2}=&\,24\pi Vr^d\,{\rm RV}\int d^dx'\log\left(r\over r'\right)\la T_{\bar z\bar z}(x)T_{zz}(x')\ra\,,
\label{eq:chiral}
\fe
which are equivalent to \eqref{Bkappa}.  

\section{Sum Rules for Defect Anomalies}\label{sec:spectrum}

The $\mH^3\times{\rm S}^{d-3}$ conformal frame exposes a simple sum rule for the defect anomalies when we perform the defect conformal block expansion of the integral formula~\eqref{Bkappa}. At $\sigma\neq0$ the defect OPE expansion converges at every $z'$\footnote{The expansion parameter is $e^{-\eta}$, with $\eta$ the geodesic distance on $\mH^3$, and it is bounded by $e^{-|\sigma|}$ uniformly in $z'$. At $\sigma=0$ the convergence is not uniform as $z'\to z$, where $\eta\to0$, so a contact term at $x'=x$ is not captured by the sum over primaries.}, so $F_\pm$ is a sum over defect conformal families, each entering through its two-point function with $T$. The average in~\eqref{FH3} keeps only transverse scalars, and the integral over $z'$ in~\eqref{Fpm} removes descendants. Conformal symmetry fixes the two-point function up to the bulk-defect OPE coefficients $c_{T\hat O}$ \cite{Billo:2016cpy,Herzog:2020bqw}: for longitudinal spin $\ell=0,1,2$, the only values that can be exchanged,
\ie
\la T_{zz}(x)\hat O_{c_1\dots c_\ell}(z')\ra={c_{T\hat O}\,\hat\Xi_z^{2-\ell}\,\hat I_{zc_1}\dots\hat I_{zc_\ell}\over r^{d-\hat\Delta}s^{2\hat\Delta}}\,,
\label{bulkdefect}
\fe
where $s^2=r^2+|z-z'|^2$, $\hat\Xi_z=r(\bar z-\bar z')/s^2$ and $\hat I_{ab}=\D_{ab}-2(z-z')_a(z-z')_b/s^2$. For $\ell<2$ this is a total derivative in $z$ and integrates to zero over $z'$. For $\ell=2$ only $\hat I_{z\bar z}=r^2/2s^2$ enters. Let $p^+_{\hat O}=c_{T\hat O}^2$ and similarly $p^-_{\hat O}$ for the bulk-defect OPE of $T_{\bar z\bar z}$. In $F_{zz\bar z\bar z}$ the exchanged defect operator is created by whichever of $T_{zz}(x)$ and $T_{\bar z\bar z}(x')$ is closer to $\Sigma$, so the weight is $p^+$ at $\sigma>0$ and $p^-$ at $\sigma<0$, and the reverse in $F_{\bar z\bar zzz}$. Hence
\ie
F_\pm(\sigma)=\sum_{\hat O}{\pi(p^+_{\hat O}\pm p^-_{\hat O})\over\hat\Delta+1}\times\begin{cases}e^{-\hat\Delta\sigma},&\sigma>0\,,\\ \pm e^{(\hat\Delta-2)\sigma}\,,&\sigma<0\,.\end{cases}
\label{sexp}
\fe
The decay at $\sigma\to-\infty$ needs $\hat\Delta>2$, which follows from unitarity since $\ell=2$ requires $\hat\Delta\ge2$, and at $\hat\Delta=2$ a conserved stress tensor on $\Sigma$ decouples, namely $c_{T\hat O}=0$ \footnote{The same bound makes the integrals of Section~\ref{sec:transport} converge. Near $\Sigma$ a primary gives $\int dr\,r^{\hat\Delta-3}$ as $r\to0$. At large distance the vector field~\eqref{xi} grows polynomially, and cutting it off at $|x'|\sim L$ leaves a shell of order $L^{2-\hat\Delta}$, which vanishes as $L\to\infty$.}.

Reflection positivity gives $p^\pm_{\hat O}\ge0$. However, $G$ is the subtracted correlator $\la TT\ra\equiv \la TT\ra_\Sigma-\la TT\ra_0$, and the part $\la TT\ra_0$ with the trivial defect has its own expansion on $\Sigma$, over the transverse derivatives $\pa_\perp^{2k}T$ of the bulk stress tensor. These enter with a minus sign, $p^+=p^-<0$, so they contribute to $B$ and cancel in $\kappa$. The first moments in~\eqref{Bsingleintegral} then give, with $p={1\over2}(p^++p^-)$ and after adding to $\kappa$ the term $6\pi V^2\int d\sigma F_-$, which vanishes by~\eqref{W2} and its $z\leftrightarrow\bar z$ image and makes its sum converge,
\ie
B=&\,24\pi^2V^2\Bigl[\sum_{\hat O}{p_{\hat O}(\hat\Delta-1)\over(\hat\Delta+1)\hat\Delta^2(\hat\Delta-2)^2}-{d-1\over d-2}A_{d-2}\Bigr]\,,\\
\kappa=&\,24\pi^2V^2\sum_{\hat O}{p^+_{\hat O}-p^-_{\hat O}\over(\hat\Delta+1)\hat\Delta^2(\hat\Delta-2)^2}\,.
\label{eq:sumrule}
\fe
Here $A_{d-2}=\sum\lambda_{TTO}a_O$, over bulk scalars of dimension $d-2$ with $\la O\ra=a_Or^{2-d}$, is a contact term at $\sigma=0$; the pseudoscalar contact term for $\kappa$, possible only in $d=3$~\cite{Costa:2011mg}, is already included in the sum. So $\kappa$ measures the chiral asymmetry of the spin-$2$ spectrum.\footnote{
A decoupled CFT on $\Sigma$ adds ${1\over2}(\hat c_L+\hat c_R)$ to $B$ and $\hat c_L-\hat c_R$ to $\kappa$. Its stress tensor $t_{ab}$ has $\hat\Delta=2$ and does not enter $F_\pm$, but the shell at $|x'|\sim L$ no longer vanishes as $L\to\infty$ and leaves the flux $
\hat c_L=4\pi\,{\rm Im}\oint_{|z|\to\infty}\!dz\,z^3\la t_{zz}(z)\,t_{zz}(0)\ra$
and its mirror image $\hat c_R$. The same holds for a gapped bulk, where $t_{ab}=\int d^{d-2}y\,T_{ab}$ is conserved and the defect is a 2d CFT at long distances.}

When $\la TT\ra$ is known in position space, as in Section~\ref{sec:examples}, it is more useful to reduce~\eqref{Bsingleintegral} to an integral over one variable. Conformal symmetry fixes the dependence on the position along $\Sigma$ through two cross-ratios, $\cosh\eta=(|z-z'|^2+r^2+r'^2)/2rr'$ and $\cos\phi=y\cdot y'/rr'$, with $\eta$ the geodesic distance on $\mH^3$ and $\phi$ the angle on ${\rm S}^{d-3}$. At fixed cross-ratios the structures surviving on $T_{zz}$ and $T_{\bar z\bar z}$ are polynomials in $\chi\equiv|z-z'|^2/(rr')$: three parity-even ones~\cite{Herzog:2020bqw} and, since the Levi-Civita tensor on $\mH^3$ gives one pseudoscalar proportional to $\sinh\sigma$, two parity-odd ones,
\ie
(rr')^dG_{zz\bar z\bar z}=\sum_{n=0}^2\chi^n f_n+\sinh\sigma\sum_{n=0}^1\chi^n g_n\,,
\label{TTstructures}
\fe
and the $g_n$ change sign in $G_{\bar z\bar z zz}$. At fixed $\eta$ the integral in~\eqref{Bsingleintegral} runs over $|\sigma|<\eta$ with $\chi=2(\cosh\eta-\cosh\sigma)$, and the conservation of the ${\rm S}^{d-3}$ average of $T$ on $\mH^3$ removes $f_2$ and $g_1$.

With bars denoting the average over ${\rm S}^{d-3}$, $\bar f_n(\eta)\equiv V^{-2}\int d\Omega\,d\Omega'\,f_n$, we find after some algebra
\ie
B=&\,16\pi^2V^2\,{\rm RV}\int_0^\infty\! d\eta\, K(\eta)\big(\sinh\eta\,\bar f_1-\pa_\eta\bar f_0\big)\\
&-24\pi^2V^2\,{d-1\over d-2}\,A_{d-2}\,,\\
\kappa=&\,32\pi^2V^2\,{\rm RV}\int_0^\infty\! d\eta\, K(\eta)\sinh\eta\,\bar g_0\,,
\label{Bkappaeta}
\fe
with the positive kernel $K(\eta)=(2\cosh^2\eta+1)\eta-3\cosh\eta\sinh\eta$. The first term of $B$ vanishes on longitudinal-spin-$0$ exchange, which contributes only at $\eta=0$, through $A_{d-2}={4\over3}\lim_{\eta\to0}\eta^5\bar f_0$; for a boundary in $d=3$, where conservation allows such a primary only at $\hat\Delta=3$~\cite{Herzog:2020bqw}, namely the displacement, this is the term $-d_1$ of Section~\ref{sec:3d}.

Finally, the formulas above hold with or without parity of the bulk CFT. Without a defect, $\la TT\ra$ has no parity-odd part at separated points for $d\geq 3$ and in $d=3$ a parity-odd term is only a contact term~\cite{Closset:2012vp}. So $F_-$ and the $\bar g_n$ come from the defect alone, and the results for $b$ in Section~\ref{sec:3d} hold for any bulk.

Bulk parity matters instead for whether $\kappa$ can be nonzero. Any map that exchanges $z$ and $\bar z$ takes a defect to one with the same $B$ and opposite $\kappa$. A parity-invariant bulk has such a map, longitudinal parity, and a defect invariant under it has $\kappa=0$. Any bulk has another one, longitudinal combined with transverse parity, which is a rotation by $\pi$ in a plane with one tangent and one normal direction. For a $U(1)$ monodromy it sends $\alpha\to-\alpha$, so $\kappa$ is odd in $\alpha$; it vanishes if the $U(1)$ commutes with parity, as for the complex scalar and the Dirac fermion of Section~\ref{sec:monodromy}, but not for the Weyl fermion, where $\kappa=\alpha$.

\section{Applications}\label{sec:examples}

\subsection{Boundaries in 3d and Bound on $b$}\label{sec:3d}

For $d=3$, the Weyl tensor vanishes and $\la T\ra=0$, so $B=b$. Earlier work fixed $d_1$ by $c_D$~\cite{Herzog:2017kkj,Jensen:2018rxu} and found further sum rules for the displacement operator~\cite{Herzog:2021spv}. 
Our identity \eqref{Bkappaeta} now gives $b$ and $\kappa$ exactly in terms of correlation function data, and it bounds $b$ from below by $c_D$. By the folding trick it suffices to study conformal boundaries, and for simplicity we assume there are no decoupled gapless degrees of freedom on the boundary. Since our definition of $\la TT\ra$ involves subtracting the trivial defect contribution, to apply~\eqref{Bkappaeta} to a boundary $|\cB\ra$ we consider the factorized interface $|\cB\ra\la\cB|$, with $\la\cB|$ the image of $|\cB\ra$ under transverse parity (or pair with $\la\overline\cB|$ by CPT if the bulk violates parity), which has $B=2b$. In the following we will study explicitly the sum rule of Section~\ref{sec:spectrum} in this setting and derive the universal bound on the Weyl anomalies in~\eqref{Weylanomaly},
\ie 
b>-d_1=-{3\pi^2\over4} c_D\,.
\label{bbound}
\fe
A decoupled CFT on the boundary would only raise $b$.

Conservation fixes the parity-even part of the boundary $\la TT\ra$ by a single function $\alpha(v)$ of the cross-ratio $v$ \cite{McAvity:1995zd,Herzog:2017xha,Herzog:2017kkj,Herzog:2020bqw},
\ie
\la T_{yy}(y)T_{yy}(y')\ra={\alpha(v)\over|y-y'|^6}\,,\quad v={|y-y'|\over y+y'}\,,
\label{TTkinem}
\fe
where the insertions are restricted to $z=z'=0$ with $y'>y$. 
Then \eqref{Bkappaeta} reduces after some algebra to
\ie
b=3\pi^2\!\int_0^1\!\!dv\,w(v)\Bigl[\alpha(v)-{c_D\over2}(1+v^6)\Bigr]\,,
\label{eq:b3d}
\fe
with $c_D=\alpha(1)$~\cite{Herzog:2017kkj} and
\ie
w(v)={3v^4-2v^2+3\over12v^4(1-v^2)^2}-{1+v^2\over4v^5}\,{\rm arctanh}\,v\,.
\fe
The subtraction, a multiple of the trivial-interface correlator $1+v^6$, which integrates to zero under the RV regulator, makes the integral converge at $v=1$.  This corrects the previous conjecture of~\cite{Herzog:2017kkj} for $b$ in terms of $\la TT\ra$.
In \cite{Prochazka:2018bpb}, the relation of the boundary anomalies to $\la TT\ra$ was studied in momentum space and we have checked that the Fourier transform of the relation there is consistent with our result.

The boundary-channel expansion
\ie 
\alpha=c_D\cD(v)+\sum p_{\hat O}\cA_{\hat\Delta}(v)\,,
\fe 
runs over the displacement and the spin-$2$ primaries \footnote{One might expect the displacement not to contribute, since it enters $T_{zz}$ only through the descendant $\pa_z^2D$, which integrates to zero over $z'$ at $\sigma\neq0$. It contributes only at $x'=x$ (i.e. $\sigma=0$ and $z'=z$) and $\eta=0$, through its portion of the contact term $A_1$ in~\eqref{Bkappaeta}, where its block has $\bar f_0\sim c_D\eta^{-5}$; if no bulk scalar of dimension $1$ has a one-point function, $A_1=0$ and this singularity cancels against the other blocks. In $d=3$ the ${\rm S}^0$ average keeps it, since $T_{yy}(0^+)+T_{yy}(0^-)$ is even under transverse parity, while for $d\ge4$ the transverse rotation symmetry removes this contribution.} with the $p_{\hat O}\ge0$ of~\eqref{sexp} and
\ie 
&\cD={1\over 8}(3v+2v^3+3v^5)\,,~
\\
&\cA_{\hat\Delta}={2^{6-2\hat\Delta}\,v(1-v^2)^{\hat\Delta-1}\over(\hat\Delta-1)(\hat\Delta-2)}\\
&\qquad\times{}_2F_1(\hat\Delta-3,\hat\Delta-{1\over 2};2\hat\Delta-1;1-v^2)\,.
\fe 
Inserting this in~\eqref{eq:b3d}, the displacement is the only negative contribution,
\ie
b=-{3\pi^2\over4}c_D+24\pi^2\sum_{\hat O}{p_{\hat O}\over(\hat\Delta^2-1)\hat\Delta^2(\hat\Delta-2)^2}\,,
\label{eq:bound}
\fe
which gives~\eqref{bbound}. 

Note that the weight in the sum rule differs from~\eqref{eq:sumrule} because each block carries its own part of the contact term $A_1$. The inequality is strict because every block vanishes at $v=0$, while $\alpha(0)={2\over3}C_T>0$ with $C_T$ the norm of the bulk stress tensor, so spin-$2$ primaries must be present.

An interface is a boundary of the folded theory, so~\eqref{bbound} holds for it with $c_D$ replaced by the norm of $T_{yy}(0^+)+T_{yy}(0^-)$, which is not the interface displacement unless the interface factorizes. For $d\ge4$ no such bound follows: the displacement drops out of the ${\rm S}^{d-3}$ average, and the sums over primaries for the defect and the trivial defect diverge separately, so positivity does not bound $B$.

The function $\alpha$ cannot give $\kappa$, since $\la T_{yy}T_{yy}\ra$ is parity even. Instead, the parity-odd part of $\la TT\ra$ is fixed by a second function at the same points,
\ie
\tilde\alpha(v)=|y-y'|^6\bigl(\la T_{zz}(y)T_{\bar z\bar z}(y')\ra-\la T_{\bar z\bar z}(y)T_{zz}(y')\ra\bigr)\,.
\fe
By \eqref{TTstructures} and \eqref{Bkappaeta},
\ie
\kappa={\pi^2\over2}\int_0^1\!\!dv\,{(1-v^2)^2\over v^6}\,K(\eta)\,\tilde\alpha(v)\,,\quad v=\tanh{\eta\over2}\,.
\label{eq:kappa3d}
\fe
Here $\kappa$ is half that of $|\cB\ra\la\cB|$ in a parity-invariant bulk: each face alone has a $\kappa$ that shifts with the gravitational Chern--Simons contact term of the bulk~\cite{Closset:2012vp}, which cancels between the two.

We test these on free fields with conventional boundary conditions, namely a scalar with Neumann ($+$) or Dirichlet ($-$) conditions and a Dirac fermion with $\gamma^3\psi=\pm\psi$, and at large $N$ on the critical $O(N)$ model with ordinary or normal boundary conditions and on the normal boundary $|B_1\ra$ of the Gross--Neveu model with $N$ two-component Dirac fermions, of mass ${1\over2}$ on $\mH^3$ (see Appendix~\ref{app:TT}). With $\alpha_{\rm ord}$ in~\eqref{alphaord}~\cite{McAvity:1995zd} and $\tilde\alpha=0$ where not shown,
\ie
{\rm scalar:}&\quad\alpha=\alpha(0)(1+v^6)\pm C_Tv(1-v^2)^2\,,\\
{\rm Dirac:}&\quad\alpha={c_D\over2}(1+v^6)\,,~\tilde\alpha={3v^2(1+v^2)\over32\pi^2}\,,\\
{\rm ordinary:}&\quad\alpha=\alpha_{\rm ord}\,,\\
{\rm normal:}&\quad\alpha=9\alpha_{\rm ord}-{N(1-v^2)^3\over2\pi^2}\,,\\
|B_1\ra:&\quad\alpha=6\alpha_{\rm ord}-{N(1-v^2)^3\over4\pi^2}\,,\\
&\quad\tilde\alpha=-{Nv(1-v^4)\over8\pi^2}\,.
\label{eq:examples}
\fe
Table~\ref{tab:3d} collects the values that follow from~\eqref{eq:b3d} and~\eqref{eq:kappa3d}. For free fields, $b$ and $d_1$ agree with~\cite{Jensen:2018rxu}, and at large $N$, $b$ agrees with the free energies of~\cite{Giombi:2020rmc}, \cite{Krishnan:2023cff} and~\cite{Giombi:2021cnr}. Only the Dirac fermion and $|B_1\ra$ break longitudinal parity. For the former, $\kappa$ agrees with direct integration of~\eqref{eq:chiral} and its image under longitudinal parity on $|\cB\ra\la\cB|$, and with anomaly matching: the strip between $\la\cB|$ and $|\cB\ra$ has a two-dimensional Weyl fermion as its zero mode, so $|\cB\ra\la\cB|$ has $\kappa=1$. For the latter, $\kappa$ agrees with~\cite{Diatlyk:2026gab}, and $|\overline{B_1}\ra$ has the opposite sign. All of them satisfy our bound on Weyl anomaly~\eqref{bbound}, and the normal $O(N)$ boundary comes close to saturating it.

\begin{table}[t]
\begin{ruledtabular}
\begin{tabular}{lcccc}
Boundary & $b$ & $d_1$ & $\kappa$ & $b/d_1$\\
\hline
Scalar: Neumann & $1/16$ & $3/32$ & $0$ & $2/3$\\
Scalar: Dirichlet & $-1/16$ & $3/32$ & $0$ & $-2/3$\\
Dirac: $\gamma^3\psi=\pm\psi$ & $0$ & $3/16$ & $\pm 1/2$ & $0$\\
$O(N)$: ordinary & $0$ & $3\pi^2N/512$ & $0$ & $0$\\
$O(N)$: normal & $-N/2$ & $27\pi^2N/512$ & $0$ & $-256/27\pi^2$\\
Gross-Neveu: $|B_1\ra$ & $-N/4$ & $9\pi^2N/256$ & $-N/2$ & $-64/9\pi^2$
\end{tabular}
\end{ruledtabular}
\caption{Boundary anomalies in 3d CFTs from stress tensor correlators \eqref{eq:examples}. All satisfy the bound $b/d_1>-1$ of~\eqref{bbound}.}
\label{tab:3d}
\end{table}

\subsection{Monodromy Defects in 4d Free Theories}\label{sec:monodromy}

Free fields in 4d with a flavor monodromy around $\Sigma$ are the simplest nontrivial surface defects~\cite{Lauria:2020emq,Giombi:2021uae,Bianchi:2021snj}. In the hyperbolic frame~\eqref{Weyl} they reduce to massive free fields on $\mH^3$, and their anomalies become simple sums over these fields.

Let us consider a complex scalar $\Phi$ or a Weyl fermion $\Psi$ with monodromy $e^{2\pi i\alpha}$ around $\Sigma$~\cite{Giombi:2021uae,Bianchi:2021snj}. On $\mH^3\times{\rm S}^1$, with $\theta$ the angle on ${\rm S}^1$, the monodromy fixes the transverse spins $s$ of the Fourier modes,
\ie
\Phi=\!\!\sum_{s\in\mZ+\alpha}\!{e^{is\theta}\phi_s\over\sqrt{2\pi}\,r}\,,\quad\Psi=\!\!\sum_{s\in\mZ+{1\over2}+\alpha}\!{e^{is\theta}\psi_s\over\sqrt{2\pi}\,r^{3/2}}\,.
\label{modes}
\fe
The $\mH^3$ mode $\phi_s$ is a scalar of mass squared $m^2=s^2-1$ and $\psi_s$ is a two-component spinor of mass $m=s$, and the corresponding defect operators have dimension $\hat\Delta=1+|s|$.

Since $T$ is bilinear in the fields, $\la TT\ra$ can be evaluated by Wick contraction via a product of two $\mH^3$ propagators~\cite{Diatlyk:2026gab}, and the average over ${\rm S}^1$ in~\eqref{FH3} sets their transverse spins equal. Each mode therefore contributes separately to~\eqref{Bkappaeta}, where its integrand is a total derivative in $\eta$, so only the boundary term at $\eta=0$ remains. A scalar mode contributes $-|s|$ to the integral in $B$, and a fermion mode contributes $-{1\over2}|s|$ to $B$ and $-{1\over2}{\rm sgn}(s)$ to $\kappa$, with the sign of $s$ setting the 2d chirality of its defect operator. These sums are regulated by the factor $e^{-2|s|\eta}$ that each mode carries, and we subtract their values at $\alpha=0$, where the correlator is that of the trivial defect. With $\tau(\A)\equiv \alpha(1-\alpha)$, the relevant sums are
\ie
\sum_{\mZ+\alpha}|s|=\tau(\A)\,,~\sum_{\mZ+{1\over2}+\alpha}|s|=-\alpha^2\,,~\sum_{\mZ+{1\over2}+\alpha}{\rm sgn}(s)=-2\alpha\,.
\label{spectralsums}
\fe
For the scalar, the integral in~\eqref{Bkappaeta} is therefore $-\tau(\A)$, and the contact term from $\Phi^\dagger\Phi$, of dimension $d-2=2$, gives $A_2=\tau(\A)$. The integral does not see the improvement term of $T$, while the contact term is proportional to it, and at conformal coupling the two cancel. In sum for a free scalar monodromy defect,
\ie 
B_{\rm scalar}(\A)=\kappa_{\rm scalar}(\A)=0\,.
\fe
The Weyl fermion with $|\alpha|<{1\over2}$ has no scalar operator of dimension $2$, so there is no contact term, and
\ie
B_{\rm fermion}(\A) ={\alpha^2\over2}\,,\quad \kappa_{\rm fermion}=\alpha\,,
\label{eq:Weyl}
\fe
where $\kappa$ flips sign for the opposite chirality. 

The same propagators at coincident points give $h$. Since $d_2=36\pi^2h$ by~\eqref{d1d2} and $b=B+12\pi^2h$ by~\eqref{Banom}, we obtain, with no curved background,
\ie
{\rm scalar:}&\quad b={1\over2}\tau(\A)^2\,,~d_2={3\over2}\tau(\A)^2\,,\\
{\rm Dirac:}&\quad b=2\alpha^2-\alpha^4\,,~d_2=3\alpha^2(1-\alpha^2)\,,
\fe
in agreement with~\cite{Giombi:2021uae,Bianchi:2021snj}.

\section*{Acknowledgments}
ZS is supported by the U.S. Department of Energy grant DE-SC0009988 and the Sivian Fund. The work of YW is supported in part by the BSF Grant 2024187, the NSF Grant 2609962 and by the Ambrose Monell Foundation and the Friends of the Institute for Advanced Study.

AI statement: we acknowledge the assistance of ChatGPT and Claude in checking various integrals and improving the exposition.

\clearpage
\onecolumngrid

\appendix
\section{Stress-tensor two-point functions at large \texorpdfstring{$N$}{N}}
\label{app:TT}

In this Appendix, we compute $\alpha(v)$ at large $N$ for the normal boundary
of the $3d$ Gross-Neveu (GN) and critical $O(N)$ models, using the hyperbolic space setup. We start with a brief review of the $\mH^3$ geometry.
Let $x^\mu=(\mathbf x,y)$, $y>0$ be the Poincar\'e coordinates of $\mH^3$ of unit radius, with the conformally flat metric $\mathrm d s^2=\frac{\mathrm d y^2+\mathrm d \mathbf x^2}{y^2}$. $X^A(x)$ denotes the embedding space coordinates of $x$, satisfying $X^AX_A=-1$. The embedding space metric is $\eta_{AB}=(-1,1,1,1)$.  Given two points $X_1$ and $X_2$, we define the AdS invariant 
\begin{align}
q\equiv-X_1\cdot X_2
&=\frac{y_1^2+y_2^2+|\mathbf x_1-\mathbf x_2|^2}{2y_1y_2}~.
\label{eq:geometry}
\end{align}
We also use the notation $\sigma=\cosh \eta$ and $v=\sqrt{\frac{q-1}{q+1}}$.
For the special configuration $\mathbf x_1=\mathbf x_2=0$, the function $\alpha(v)$ of~\eqref{TTkinem} is given in this Weyl frame by
\begin{equation}
\vev{T_{yy}(y_1, 0)T_{yy}(y_2, 0)}_c
=\frac{y_1y_2}{(y_1-y_2)^6}\,\alpha(v)~.
\label{eq:alpha}
\end{equation}
Here we only consider the traceless representative of $T_{\mu\nu}$ defined by metric variation because the contact terms do not impact two point functions at separated points.

For later convenience, we  introduce the following traceless differential operator
\begin{equation}
D_{\mu\nu}=\nabla_\mu\nabla_\nu-\frac13g_{\mu\nu}\lap .
\label{eq:D}
\end{equation}
For an arbitrary AdS invariant bifunction $f(X_1, X_2) = f(q)$, the action of $D_{\mu\nu}$ takes the form
\begin{equation}
D_{\mu\nu}f(q)=
\left[\nabla_\mu q\nabla_\nu q 
-\frac13g_{\mu\nu}(q^2-1)\right]f''(q)\,.
\label{eq:Dscalar}
\end{equation}
Here, $x^\mu$ are local coordinates of either $X_1$ or $X_2$, and primes denote $q$ derivatives. To derive this formula, we have used $\nabla_\mu X_A \nabla^\mu X_B = \eta_{AB}+X_A X_B$ and $\nabla_\mu\nabla_\nu X_A\propto g_{\mu\nu}$~.
Another useful formula is 
\begin{equation}
D^{(1)}_{yy}D^{(2)}_{yy}f
=\frac{4}{9\, y_1^2y_2^2}\left[
(q^2-1)^2f^{(4)}
+4q(q^2-1)f^{(3)}
+(2q^2+1)f''\right]\,,
\label{eq:radial}
\end{equation}
where $D_{yy}^{(i)}$ acts on the point $X_i$.

Next, we review the main ingredients of harmonic analysis in hyperbolic space. The spectral density of $f(X_1, X_2)$ is defined by 
\begin{equation}
f(X_1, X_2)=\int_{\mathbb R}\mathrm d\nu\,\rho_f(\nu)\Omega_\nu(X_1, X_2)~,
\label{eq:spectral}
\end{equation}
where $\Omega_\nu$ is  the harmonic function. In the $\eta$ variable, it can be expressed as $\Omega_\nu(\eta)=\frac{\nu\sin(\nu \eta)}{4\pi^2\sinh \eta}$. Given the function $f$, we can find its spectral density using the inversion formula
\begin{align}
  \rho_f(\nu)=\frac{4\pi}{\nu}
\int_0^\infty \mathrm d \eta\,\sinh \eta\sin(\nu \eta)f(\eta)~.
\end{align}
For instance, a free scalar of boundary dimension $\hat\Delta=1+a$ has the propagator 
\begin{gather}
    G_a(\eta)=\frac{e^{-a\eta}}{4\pi\sinh \eta}\,,
    \label{GpropDimA}
\end{gather}
and the corresponding spectral density is $\rho_a(\nu) =\frac{1}{\nu^2+a^2}$.

The harmonic function satisfies the orthogonality condition
\begin{align}
    \int_{X_2}\Omega_\nu(X_1, X_2)\Omega_{\bar\nu} (X_2,X_3)=\frac{\delta(\nu-\bar\nu)+\delta(\nu+\bar\nu)}{2}\Omega_\nu(X_1,X_3)~,
\end{align}
where $\int_X$ is the shorthand notation for the integral on $\mathbb{H}_3$.
It then follows that the spectral density of $f*g(X_1, X_3)\equiv\int_{X_2}f(X_1, X_2) g(X_2, X_3) $ is simply the product of $\rho_f(\nu)$ and $\rho_g(\nu)$. This is the crucial property for our derivation below.

Both results below are expressed through the ordinary boundary of the critical $O(N)$ model, where all $\phi^I$ have boundary dimension $\hat\Delta=1$, for which at leading order~\cite{McAvity:1995zd}
\begin{align}
\alpha_{\rm ord}(v)={}&\frac{Nv^3}{128}\left(1+\frac{3(1-v^2)^2}{8v^2}\right)+\frac{N}{16\pi^2}(1-v^2)^3\left[1-\frac19\,{}_3F_2\!\left(\begin{matrix}\frac12,\;1,\;3\\[1pt]\frac52,\;\frac52\end{matrix}\;\bigg|\;-\frac{(1-v^2)^2}{4v^2}\right)\right]\,.
\label{alphaord}
\end{align}

\subsection{$B_1$ boundary universality class of the Gross-Neveu model}
\label{App:alphaB1}

At the critical point, the large $N$ Gross-Neveu model in $\mathbb{H}_3$ is described by the action 
\begin{equation}
S_{\GN}=-\int_X \bar\psi^I(\slashed\nabla+\phi)\psi^I, 
\label{eq:GNaction}
\end{equation}
where $\psi^I$ are $N$ 2-component Dirac spinors.
The $B_1$ boundary phase corresponds to the constant saddle $\langle \phi\rangle = \frac{1}{2}$ in the large $N$ limit and choosing the standard boundary condition for the fermions \cite{Giombi:2021cnr}. 
So, the fermion propagator at the leading order of large $N$ expansion is \cite{Giombi:2021cnr}
\begin{equation}
S_\star(x_1,x_2)
=-\frac{(q+1)\gamma^a x_{12}^a
-(q-1)\gamma^3\gamma^a\bar x_{12}^a}
{8\pi\sqrt{y_1y_2}\,(q^2-1)^{3/2}},
\label{eq:spinor}
\end{equation}
where $x_{12}=x_1-x_2$, and $\bar x_{12}=(\mathbf x_1-\mathbf x_2,-y_1-y_2)$ as well as $\gamma^a$ satisfy $\{\gamma^a,\gamma^b\}=2\delta^{ab}$.
It obeys $(\slashed\nabla+\langle \phi\rangle)S_\star(X_1,X_2) = - \delta(X_1,X_2)$.
With the action convention in \eqref{eq:GNaction}, the traceless
stress energy tensor is
\begin{equation}
T_{\mu\nu}=-\frac14\left[
\bar\psi^I\gamma_\mu\overleftrightarrow\nabla_\nu\psi^I+
\bar\psi^I\gamma_\nu\overleftrightarrow\nabla_\mu\psi^I
-\frac23g_{\mu\nu}
\bar\psi^I\gamma^\alpha\overleftrightarrow\nabla_\alpha\psi^I\right].
\label{eq:TF}
\end{equation}

The fermion bubble diagram in the Gaussian theory, with all flavor indices contracted, takes the form
\begin{equation}
K(x,x')\equiv\vev{\bar\psi\psi(x)\bar\psi\psi(x')}_{0}
=-N\operatorname{tr}[S_\star(x,x')S_\star(x',x)]
=\frac{N}{8\pi^2\sinh^4\eta}.
\label{eq:KF}
\end{equation}
Its spectral density has been evaluated in \cite{Diatlyk:2026gab}:
\begin{equation}
\rho_K(\nu)=-\frac{N(\nu^2+1)}{8\nu}\tanh\frac{\pi\nu}{2}.
\label{eq:KFdist}
\end{equation}
The $\delta\phi$ correlation function is the inverse of $-K$ at large $N$. Thus, the corresponding spectral density is
\begin{equation}
\rho_{\delta\phi}(\nu)
=-\frac1{\rho_K(\nu)}
=\frac{8\nu\coth(\pi\nu/2)}{N(\nu^2+1)}.
\label{eq:GsF}
\end{equation}
In our convention, the $\delta\phi$ propagator starts at order $1/N$.

Next, we proceed to compute the $TT$ two-point functions. At order $N$, there are two contributions. The first contribution arises from Wick contractions of fermions in the Gaussian theory with the propagator \eqref{eq:spinor}.
Carrying out this lengthy calculation yields
\begin{equation}
\alpha^f_{B_1}(v)
=\frac{N}{72\pi^2}(1-v^2)(9v^4-2v^2+9).
\label{eq:Floop}
\end{equation}
The second contribution corresponds to exchanging a single $\phi$ field. The first step is to determine $\vev{T_{\mu\nu}(x)\phi(x')}_c$ at order $N^0$. We can fix it  without doing additional calculations following the logic of \cite{McAvity:1995zd}. The tracelessness of $T_{\mu\nu}$ implies that this two-point function can be realized by acting with $D_{\mu\nu}$ on an AdS invariant scalar function. The conservation of $T_{\mu\nu}$ further  requires the scalar function to be proportional to $G_2$, the scalar bulk two-point function of boundary dimension $3$, defined by \eqref{GpropDimA}. The remaining overall coefficient follows from the bulk OPE of $T_{\mu\nu}$ and $\phi$, as well as the saddle point value $\langle\phi\rangle=1/2$. Altogether, we obtain
\begin{equation}
\vev{T_{\mu\nu}(x)\phi(x')}_c=\vev{T_{\mu\nu}(x)\delta\phi(x')}_c
=-\frac{1}{4}D_{\mu\nu}^{(x)}G_2(x,x').
\label{eq:TFs}
\end{equation}
We also need to find the two-point function
$t_{\mu\nu}=\vev{T_{\mu\nu}\bar\psi\psi}_{0,c}$, evaluated in the Gaussian theory. It is equivalent to $\vev{T_{\mu\nu}(x)\delta\phi(x')}_c$ with the $\delta\phi$ propagator removed. This can be achieved by using the fermion bubble function $K$, explicitly
\begin{equation}
t_{\mu\nu}=-\vev{T_{\mu\nu}\delta\phi}_c*K .
\label{eq:amputation}
\end{equation}
Thus the $TT$ two-point function due to a single $\phi$ exchange takes the simple form
\begin{align}
\langle T_{\mu\nu} T_{\rho\sigma}\rangle_{\rm ex}
&=-\vev{T_{\mu\nu}\delta\phi}_c*K*
\vev{\delta\phi T_{\rho\sigma}}_c\notag=-\frac{1}{16}D^{(1)}_{\mu\nu}D^{(2)}_{\rho\sigma}
(G_2*K*G_2)~,
\label{eq:CFex}
\end{align}
where $G_2*K*G_2$ has spectral density
$\rho_K/(\nu^2+4)^2$. Using \eqref{eq:KFdist}, we get
\begin{equation}
\langle T_{\mu\nu} T_{\rho\sigma}\rangle_{\rm ex}
=\frac N{128}D^{(1)}_{\mu\nu}D^{(2)}_{\rho\sigma}
\int_{\mathbb R}\mathrm d\nu\,
\frac{(\nu^2+1)\tanh(\pi\nu/2)}
{\nu(\nu^2+4)^2}\,\Omega_\nu(x_1,x_2)\,.
\label{eq:GNdirectspectral}
\end{equation}
Computing the spectral integral summing up the poles leads to 
\begin{equation}
\begin{aligned}
 \int_{\mathbb R}\mathrm d\nu\,
\frac{(\nu^2+1)\tanh(\pi\nu/2)}
{\nu(\nu^2+4)^2}\,&\Omega_\nu(x_1,x_2)=\frac1{16\pi^2}\Big[
1-5q\log v-\frac{3\pi^2}{2}q\\
&+\frac{3(2q^2-1)}{\sinh \eta}
\left(\Li(v)-\Li(-v)-2\log v\,\atanh v\right)\Big]~,
\end{aligned}
\label{eq:GNintegralclosed}
\end{equation}
where $q = \frac{1+v^2}{1-v^2}$. Finally, applying $D^{(1)}_{yy}D^{(2)}_{yy}$ to the integral gives 
\begin{equation}
\begin{aligned}
\alpha^{\rm ex.}_{B_1}(v)=\frac{N}{64\pi^2}\Big[
&\frac79v^2(1-v^2)+9v^2(1+v^2)\log v\\
&+\frac32v(3v^4+2v^2+3)
\left(\Li(v)-\Li(-v)-2\log v\,\atanh v\right)\Big].
\end{aligned}
\label{eq:GNexdilog}
\end{equation}
Adding \eqref{eq:Floop} and \eqref{eq:GNexdilog} and expressing
the result in terms of the ordinary boundary function
$\alpha_{\rm ord}(v)$ in \eqref{alphaord}, we obtain
\begin{equation}
\alpha_{B_1}(v)=6\,\alpha_{\rm ord}(v)-{N(1-v^2)^3\over4\pi^2}\,,
\label{alphaB1}
\end{equation}
as quoted in Section~\ref{sec:3d}.

\subsection{Normal boundary of the $3d$ $ O(N)$ CFT}\label{App:alphanormal}
The critical point of the $O(N)$ vector model at large $N$, placed in $\mathbb{H}_3$, is described  by the action 
\begin{equation}
S=\frac12\int_X\left[
(\nabla\phi^I)^2-\frac34\phi^2+\lambda\phi^2\right],
\qquad
\lambda=\lambda_*+\delta\lambda,\qquad
\phi^N=\phi_*+\delta\phi^N .
\label{eq:ONaction}
\end{equation}
where $-\frac{3}{4}\phi^2$ is the conformally coupled mass, and $\lambda$ is the Hubbard-Stratonovich field, whose quadratic term can be dropped at the IR fixed point.
At the normal boundary fixed point, $\lambda$ sits at the saddle point $\lambda_*=\frac{3}{4}$, and $\phi^I$ also picks up a nontrivial vacuum expectation value (VEV). We rotate the $\phi$ field such that the first $N-1$ components vanish at the saddle point. The VEV of $\phi^N$ is $\phi_*^N = \sqrt{\frac{N}{4\pi}}$. At this saddle point field configuration, the transverse fields $\phi^a$, $a=1,\ldots,N-1$ are massless scalars of boundary dimension $\hat\Delta=2$, with bulk propagators being $G_1(r) = \frac{e^{-\eta}}{4\pi\sinh(\eta)}$.
Integrating out the transverse fields yields a mixed action of the $\lambda$ fluctuation, denoted by $\delta\lambda$, and the $\phi^N$ fluctuation, denoted by $\delta\phi^N$. At the quadratic level, this action reads
\begin{equation}
S_2[\delta\phi^N, \delta\lambda]=\int_X\frac12\delta\phi^N(-\lap)\delta\phi^N
+\phi^N _* \delta\phi^N\delta\lambda
-\frac N4\,\int_{X_1, X_2}\delta\lambda(X_1)\,G_1(X_1, X_2)^2\,\delta\lambda(X_2).
\end{equation}
Completing the square in $\delta\phi^N$ gives a quadratic action of $\delta\lambda$ 
\begin{equation}
S_2[\delta\lambda]=
-\frac N4\,\int_{X_1, X_2}\delta\lambda(X_1)\,G_0(X_1, X_2)^2\,\delta\lambda(X_2).
\end{equation}
The spectral density of $\delta\lambda$ is thus $\rho_{\delta\lambda}(\nu)
=-\frac2{N}\rho_{G_0^2}(\nu)^{-1}$, where
\begin{equation}
 \quad \rho_{G_0^2} (\nu)= \frac{\tanh(\pi\nu/2)}{8\nu}~.
\label{eq:normal-bubble}
\end{equation}

With all these ingredients, we can compute $\alpha(v)$. The traceless stress energy tensor reads
\begin{equation}
T_{\mu\nu}
=-\phi^ID_{\mu\nu}\phi^I+\frac38D_{\mu\nu}\phi^2 = T^\perp_{\mu\nu}+L_{\mu\nu}\delta\phi^N+\cdots .
\label{eq:Tscalar}
\end{equation}
Here $T^\perp_{\mu\nu}=-\phi^a D_{\mu\nu}\phi^a+\frac38D_{\mu\nu}(\phi^a\phi^a)$ only contains the transverse fields, $L_{\mu\nu}=-\frac{\phi_*}{4}D_{\mu\nu}$, and ``$\cdots$'' denotes terms whose contribution is subleading in $1/N$ expansion.

A direct Wick contraction  calculation yields
\begin{equation}
t_{\mu\nu}(x,x')
=\frac12\vev{T^\perp_{\mu\nu}(x)\phi^a\phi^a(x')}_{0,c}
=D_{\mu\nu}\left[-\frac{3N}{8}G_2*G_0^2+\frac{N}{16\pi}G_1\right].
\label{eq:Vpi}
\end{equation}
Other two-point functions such as $\vev{T_{\mu\nu}\delta\lambda}_c$ and $\vev{T_{\mu\nu}\delta\phi^N}_c$ can be determined using the same symmetry-based argument as in the GN case:
\begin{equation}
\vev{T_{\mu\nu}\delta\lambda}_c=-\frac34D_{\mu\nu}G_2,
\qquad
\vev{T_{\mu\nu}\delta\phi^N}_c=-\frac{\phi_*}{4}D_{\mu\nu}G_2.
\label{eq:mixed}
\end{equation}

With some manipulations, the $\langle TT\rangle$ can be brought to the following form 
\begin{align}
\langle T_{\mu\nu}(x) T_{\rho\sigma}(x')\rangle 
=\langle T^\perp_{\mu\nu}(x) T^\perp_{\rho\sigma}(x')\rangle^f
&-\int_X t_{\rho\sigma}(x',X)
\vev{T_{\mu\nu}(x)\delta\lambda(X)}_c
+L_{\rho\sigma}^{(x')}
\vev{T_{\mu\nu}(x)\delta\phi^N(x')}_c~.
\label{eq:regroup}
\end{align}
Here, $\langle T^\perp_{\mu\nu}(x) T^\perp_{\rho\sigma}(x')\rangle^f$ corresponds to Wick contraction of the transverse fields in the Gaussian theory. Its contribution to $\alpha_{\rm norm}(v)$ is found to be
\begin{equation}
\alpha_{\rm norm}^{f}(v)
=\frac{N(1-v)
(v^6-v^5+2v^4+4v^3+2v^2-v+1)}
{16\pi^2(1+v)}~.
\label{eq:pi-loop}
\end{equation}
Plugging \eqref{eq:Vpi} and \eqref{eq:mixed} into \eqref{eq:regroup}, the second term, which we call the exchange part,  becomes
\begin{equation}
\langle T_{\mu\nu}(x) T_{\rho\sigma}(x')\rangle_{\rm ex.}
=\frac34D^{(1)}_{\mu\nu}D^{(2)}_{\rho\sigma}
\left[-\frac{3N}{8}G_2*G_0^2*G_2+\frac{N}{16\pi}G_2*G_1\right].
\end{equation}
We can directly compute $G_2*G_0^2*G_2$ following the same strategy in the GN case. Alternatively, we note that this term also appears in the exchange contribution to $\langle TT\rangle$ in the presence of an ordinary boundary condition \cite{McAvity:1995zd}, and has been explicitly computed there. Therefore,
\begin{align}\label{ex}
\langle T_{\mu\nu}(x) T_{\rho\sigma}(x')\rangle_{\rm ex.}&=9\langle T_{\mu\nu}(x) T_{\rho\sigma}(x')\rangle_{\rm ex.}^{\rm ord}
+\frac{3N}{64\pi}D^{(1)}_{\mu\nu}D^{(2)}_{\rho\sigma}
G_2*G_1.\nonumber\\
&=9\langle T_{\mu\nu}(x) T_{\rho\sigma}(x')\rangle_{\rm ex.}^{\rm ord}
+\frac{N}{64\pi}D^{(1)}_{\mu\nu}D^{(2)}_{\rho\sigma}
(G_1-G_2),
\end{align}
where we have used the simple fact $G_2*G_1=\frac{G_1-G_2}{3}$.  The last term in  \eqref{eq:regroup}, we call the longitudinal part, takes a very simple form
\begin{equation}
\langle T_{\mu\nu}(x) T_{\rho\sigma}(x')\rangle_{\rm long.}
=\frac{\phi_*^2}{16}D^{(1)}_{\mu\nu}D^{(2)}_{\rho\sigma}G_2
=\frac{N}{64\pi}D^{(1)}_{\mu\nu}D^{(2)}_{\rho\sigma}G_2 ~.
\end{equation}
It is canceled by the $-G_2$ term in \eqref{ex}. Combining \eqref{eq:pi-loop} together with the exchange and longitudinal contributions, and rewriting it in terms of $\alpha_{\rm ord}$ \eqref{alphaord}, we obtain  that
\begin{align}
    \alpha_{\rm norm}(v)=\alpha^f(v)+\alpha_{\rm ex.}(v)+\alpha_{\rm long.}(v) =9 \alpha_{\rm ord}(v)-\frac{N \left(1-v^2\right)^3}{2 \pi ^2}\,,
\label{alphanorm}
\end{align}
as quoted in Section~\ref{sec:3d}.

\bibliographystyle{apsrev4-1}
\bibliography{SurfaceDefect}

\end{document}